\documentclass{ws-procs961x669}            

\begin{document}
\title{Coherent deeply virtual Compton scattering off He nuclei}

\author{S. Fucini, M. Rinaldi and S. Scopetta$^*$}

\address{Dipartimento di Fisica e Geologia, University of Perugia and INFN, Perugia Section\\
Perugia, I-06123, Italy\\
$^*$E-mail: 
sergio.scopetta@unipg.it
}

\begin{abstract}
The status of realistic calculations of nuclear generalized parton distributions,
entering the theoretical description of coherent 
deeply virtual Compton scattering off nuclei, is reviewed for trinucleons and for $^4$He, also in view
of forthcoming measurements at the Jefferson Laboratory and at the future Electron Ion Collider.
\end{abstract}


\bodymatter

\section{Introduction}

Nuclear modifications of the nucleon parton structures,
discovered by the European Muon Collaboration 
\cite{Aubert:1983xm} several decades ago, cannot be explained by means of inclusive measurements only.
One of the possible ways out is to perform nuclear
imaging, now possible for the first time
through deeply virtual Compton scattering (DVCS) and deeply-virtual meson production, using the tool of generalized parton distributions
(GPDs) (see Refs \cite{Dupre:2015jha,Cloet:2019mql} for recent reports).
The comparison of the transverse spatial quark and gluon distributions in nuclei
or bound nucleons (to be obtained in coherent or incoherent DVCS, respectively) to
the corresponding quantities in free nucleons,
will allow ultimately a pictorial representation of the EMC effect.
The relevance of non-nucleonic degrees of freedom, 
as addressed in Ref. \cite{Berger:2001zb},
or the 
change of size for bound nucleons, will be observed. 
The most discussed sector of the EMC effect is
the valence region at intermediate $Q^2$, which
will be investigated by 
Jefferson Lab (JLab) at 12~GeV.
For the lightest nuclei, $^2$H, $^3$He, $^4$He,
sophisticated calculations of conventional effects,
although challenging, are possible.
This would allow one to distinguish them
from exotic ones, likely responsible for
the observed EMC behavior.
Without realistic benchmark calculations, 
making use of wave functions obtained as
exact solution of the Schr\"odinger equation
using realistic nucleon nucleon potentials
and three-body forces whenever appropriate,
the interpretation of experimental data is difficult.
Among few-body nuclei, in this talk we will concentrate
in three- and four-body systems.

\begin{figure}[t]
\hspace{3.cm}
\includegraphics[scale=0.35,angle=0]{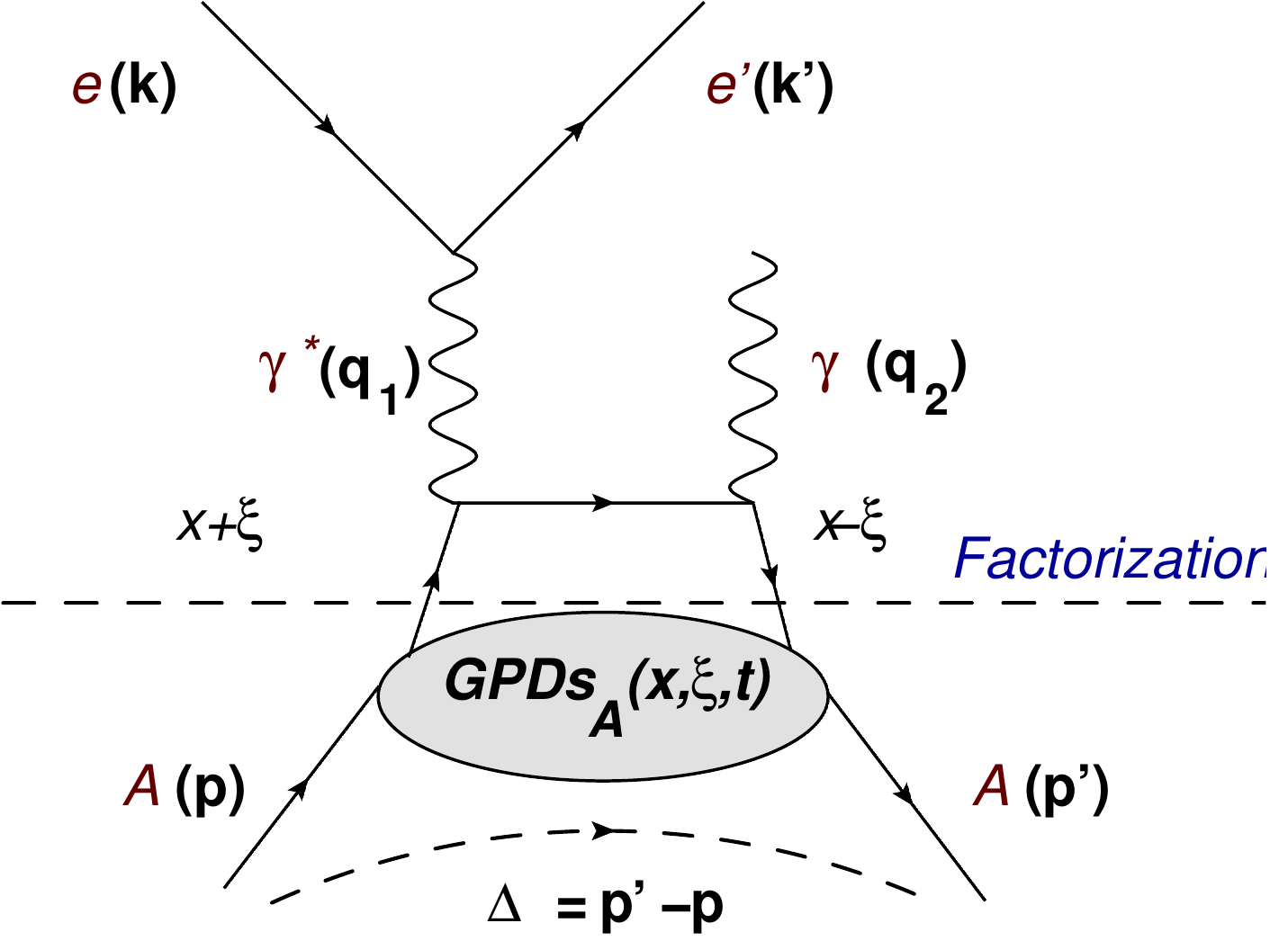}
\caption{
Coherent DVCS process off a nucleus $A$ in the handbag approximation.
}
\label{dvcsinco}
\end{figure}
\section{Coherent DVCS off $^3$He}

For the coherent channel of DVCS, the one where
the nucleus does not break up (see Fig. 1 for a
representation of the process with a generic $A$ nucleus, in the handbag approximation, i.e., with the interaction occurring on a leading quark),
due to very small cross sections,
the measurements addressed in the Introduction are very difficult.
In between the $^2$H nucleus, very interesting for its rich spin structure and for the the possible extraction of neutron information, and
$^4$He, ideal to study nuclear effects, being deeply-bound, scalar and isoscalar, with a simple
description of its spin and flavor structure,
$^3$He provides an opportunity
to study the $A$ dependence of nuclear effects, and
it could give easy access to neutron polarization properties,
due to its specific spin structure.
In addition, being isospin-$1/2$, it guarantees
that flavor dependence of nuclear effects can be studied, 
in particular if parallel measurements 
on $^3$H targets,
likely possible at the Electron Ion Collider (EIC), were performed \cite{Scopetta:2009sn}.
\begin{figure}[t]
\vskip -.9cm
\centering\includegraphics[scale=0.30]{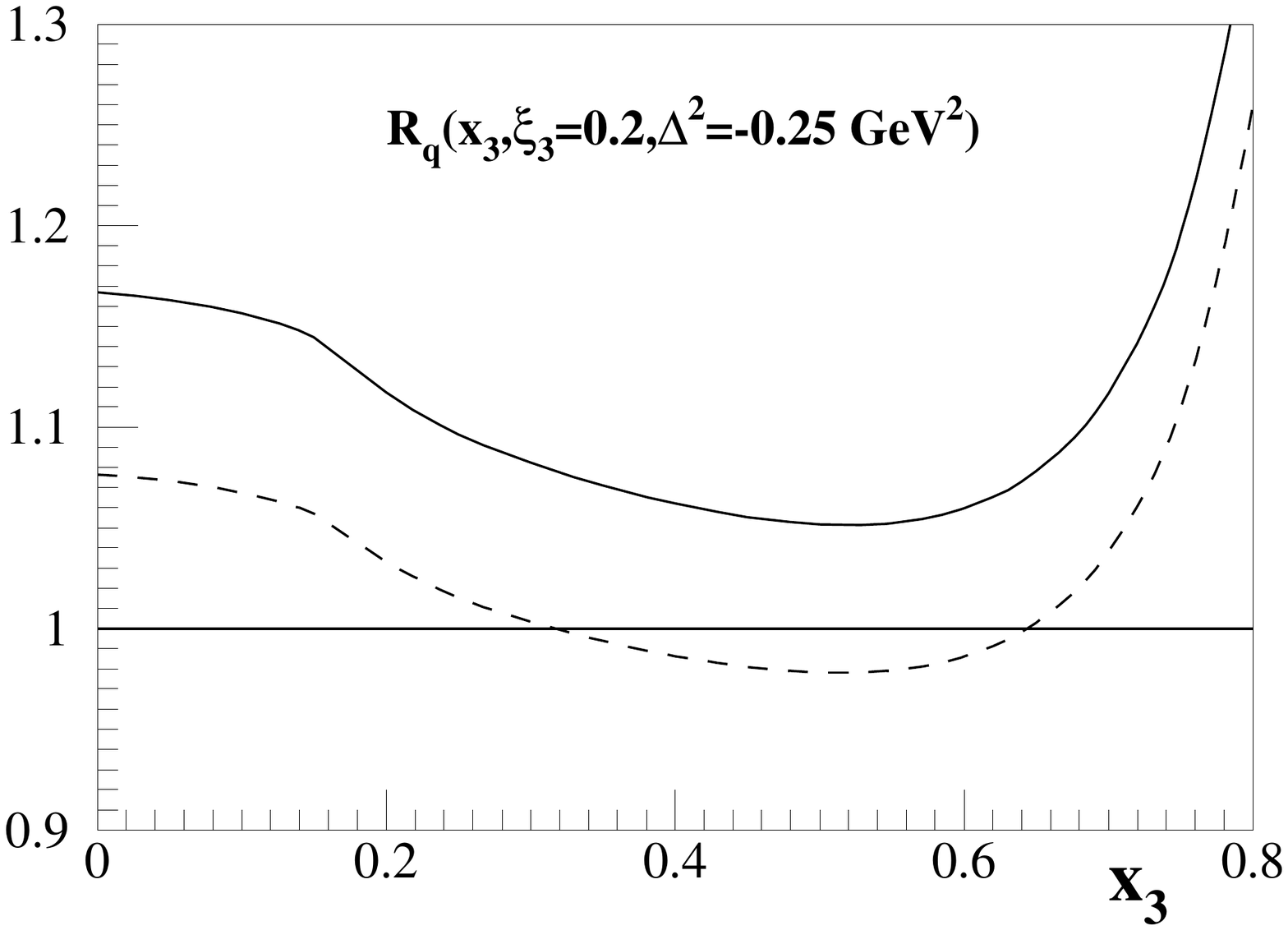}
\vskip -3.9cm
\caption{ Estimate of nuclear effects on $^3$He GPD
for the $d (u)$ flavor, full (dashed) line. The effect is given by the difference of the curves with 
respect to one.}
\label{alu}
\end{figure}
A complete realistic study of leading twist 
DVCS requires the evaluation of nuclear GPDs.
From the theoretical point of view, conventional effects for
nuclear systems are seen in a plane wave impulse approximation (IA)
analysis, i.e., with the struck quark belonging to one nucleon in the target, and disregarding possible final state interaction effects between this nucleon and the remnants).  This requires the evaluation of realistic non-diagonal spectral
functions \cite{Scopetta:2004kj}.
For $^3$He, a complete analysis using the Av18
nucleon-nucleon (NN) potential is available
\cite{Scopetta:2009sn,Rinaldi:2012pj,
Rinaldi:2012ft,Rinaldi:2014bba,Scopetta:2004kj}.
Nuclear GPDs are found to be sensitive to details of the used NN interaction.
In particular, nuclear effects are found to grow with the momentum transfer to the target, $\Delta^2$, and with the longitudinal momentum asymmetry of the process
(parametrized by the so-called skewness variable, $\xi$). In $^3$He, nuclear effects are found to be bigger for the $d$ flavor than for the $u$  one (see Fig. 2), a prediction of a realistic impulse approximation (IA), where also violations of nuclear charge symmetry are considered in the AV18 NN interaction, which could
be tested experimentally. Besides, the dependence on the excitation energy of the nuclear recoiling system in the IA description, parametrized by the so-called removal energy, is found to be bigger in nuclear GPDs than in inclusive observables.
Anyway, it is also found that close to the forward limit the information on neutron polarization can be safely extracted from 
$^3$He data and workable extraction formulae have been proposed in this sense\cite{Rinaldi:2012pj,
Rinaldi:2012ft,Rinaldi:2014bba}.
Measurements for $^3$He and $^3$H
are not planned, but could be considered
as extensions of the impressive ALERT detector
project at JLab 12
\cite{Armstrong:2017zqr,Armstrong:2017zcm}, at least in the unpolarized 
sector. Polarized measurements, which could
access neutron angular momentum information \cite{Rinaldi:2012pj,
Rinaldi:2012ft,Rinaldi:2014bba}, seem
very unlikely at JLab, due to the difficulty of arranging
a polarized target and a recoil detector in the same experimental setup, but are in principle accessible
at the EIC, where
the extension to lower
$x$ regions will be also possible \cite{Accardi:2012qut}. 

\section{Coherent DVCS off $^4$He}

Despite the difficulty of measuring coherent
DVCS off nuclei, due to small cross-sections, the first data for coherent DVCS off $^4$He
collected at JLab in the 6~GeV setup
have been published \cite{Hattawy:2017woc}.
A new impressive program is on the way at JLab,
carried on by the CLAS collaboration with the ALERT detector project 
\cite{Armstrong:2017zqr,Armstrong:2017zcm}.
\begin{figure}[t]
\centering\includegraphics[scale=0.24]{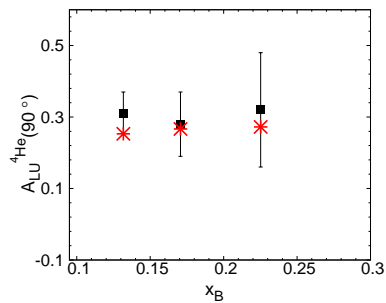}
\vskip -.5cm
\caption{ $^4$He azimuthal
(with respect to the angle $\phi$ between lepton and nuclear planes)
beam-spin asymmetry $A_{LU}(\phi)$: results of Ref. \cite{Fucini:2018gso} (red stars) compared with data
(black squares)
\cite{Hattawy:2017woc}.
}
\end{figure}
A study for DVCS off $^4$He with nuclear ingredients of the same quality of those summarized above 
for $^3$He is still missing and should be done, to update 
existing calculations,
performed long time ago\cite{Guzey:2003jh,Liuti:2005gi}. 
The evaluation of a realistic spectral functions of $^4$He,
using state-of-the-art NN potentials,
will require in particular the wave function of a nuclear three-body scattering state, 
which is a really challenging few-body problem.
An encouraging calculation has been recently
performed for coherent DVCS off $^4$He \cite{Fucini:2018gso}, with the aim to
describe the CLAS data
\cite{Hattawy:2017woc},
as a relevant intermediate
step towards a rigorous realistic evaluation.
A model of the nuclear non-diagonal spectral function, 
based on the momentum distribution
corresponding to the Av18 NN interaction
\cite{PhysRevC.67.034003}, has been used in the actual IA calculation. 
In particular the spectral function 
is exact in its ground state part (when the remnant is a bound three-body system)
and modelled in the complicated excited sector.
As a test of the procedure, typical results
are found for the nuclear form factor
and for nuclear parton distributions, in proper limits.
Nuclear GPD and the actual observable, 
the so-called,
Compton form factor
(CFF) are evaluated using
a well known GPD model to take into account the nucleonic information~\cite{Goloskokov:2011rd}. 
As can be seen in Fig.~\ref{alu}, a very good agreement is found with the data, for the so-called beam-spin asymmetry, theoretically obtained in terms of the CFF, in turn evaluated from the GPD.
One can conclude that 
a careful analysis of the reaction mechanism in terms of
basic conventional ingredients is successful and that
the present experimental accuracy does not require
the use of exotic arguments, such as dynamical off-shellness.
More refined nuclear calculations will be certainly necessary for the expected improved accuracy of the next generation of experiments 
at JLab, with the 12 GeV electron beam and high luminosity, and, above all, at the EIC.
Very recent results for the incoherent channel are reported in \cite{Fucini:2019xlc},
where an encouraging comparison with
data from JLab \cite{Hattawy:2018liu} is presented.

This work was supported in part by the STRONG-2020 project of the European Union’s Horizon 2020
research and innovation programme under grant agreement No 824093, and by
the project ``Deeply Virtual Compton Scattering off $^4$He", in the programme FRB of the University
of Perugia.
  


\end{document}